\documentclass[prl,twocolumn,groupedaddress,nopacs]{revtex4}

\usepackage{graphicx}

\begin{document}

\title{Dynamic facilitation explains democratic particle motion of
metabasin transitions}

\author{Lester O. Hedges}

\affiliation{School of Physics and Astronomy, University of Nottingham,
Nottingham, NG7 2RD, UK}

\author{Juan P. Garrahan}

\affiliation{School of Physics and Astronomy, University of Nottingham,
Nottingham, NG7 2RD, UK}

\begin{abstract}
Transitions between metabasins in supercooled liquids seem to occur
through rapid ``democratic'' collective particle rearrangements. 
Here we show that this apparent homogeneous particle motion is a direct consequence of dynamic
facilitation.  We do so by studying metabasin transitions in facilitated spin models and constrained lattice gases.  We find that metabasin transitions occur
through a sequence of locally facilitated events taking place over a
relatively short time frame.  When observed on small enough spatial
windows these events appear sudden and homogeneous.  Our results
indicate that metabasin transitions are essentially ``non-democratic''
in origin and yet another manifestation of dynamical heterogeneity in glass
formers.
\end{abstract}


\maketitle

\noindent
\section{Introduction}  
In this paper we study metabasin (MB)
\cite{Doliwa-Heuer} transitions, that is, transitions between low
energy/low activity configurations, in kinetically constrained models
(KCMs) of glass formers
\cite{Fredrickson-Andersen,Jackle,Kob-Andersen,Jackle-tlg,Ritort-Sollich}.
We follow closely the recent work of Appignanesi et al.\
\cite{Appignanesi} (see also \cite{Appignanesi2,Frechero}) 
who studied this problem in an atomistic model, a Lennard-Jones binary mixture \cite{Andersen}, by means of
molecular dynamics simulations.  In Ref.\ \cite{Appignanesi} it was found that transitions between metabasins involved relatively fast and collective rearrangements of a significant number of particles forming 
compact clusters in space.  These apparent homogeneous relaxation events were termed ``democratic'' \cite{Appignanesi}.  A natural question
is whether democratic events are distinct in nature to those associated with dynamic heterogeneity \cite{DH}. Here we show that the apparent democratic particle motion that occurs during a metabasin transition is a direct result of dynamic facilitation, an intrinsically ``non-democratic'' property of KCMs which is the origin of dynamic heterogeneity \cite{Harrowell,Garrahan-Chandler} in these systems.

\bigskip

\noindent
\section{Metabasin transitions in facilitated spin models} 
The simplest microscopic models built on the idea of dynamic facilitation 
are the so-called facilitated spin models (FSMs), such as the Fredrickson-Andersen (FA) model \cite{Fredrickson-Andersen} and its directional counterpart the East model \cite{Jackle}. In one dimension these models are described by a chain of Ising spins $n_i=\{0;1\}$, with a trivial Hamiltonian $H=\sum_i n_i$. Here $n_i=1$ represents a mobile site, or excitation, conversely $n_i=0$ represents an immobile, or jammed site. Glassiness is the result of local dynamical rules that specify the ability of a site to change state. In the FA model a site can only flip if either of its nearest neighbours are in the excited state. The transition rates are: $10 \stackrel{\epsilon}{\longrightarrow} 11$, $~11\stackrel{1}{\longrightarrow} 10$, $~01\stackrel{\epsilon}{\longrightarrow} 11$, $~11\stackrel{1}{\longrightarrow} 01$, where $\epsilon = e^{-\beta}$ and $\beta=1/T$. For the East model the last two transitions are not allowed: a spin may only flip if its nearest neighbour to the left is excited, consequently excitations propagate in an eastward direction. Relaxation is Arrhenius in the FA and super-Arrhenius in the East model \cite{Fredrickson-Andersen,Jackle,Sollich-Evans,Ritort-Sollich,Garrahan-Chandler}.

\begin{figure*}[t]
  \centering
  \includegraphics[width=2.1\columnwidth]{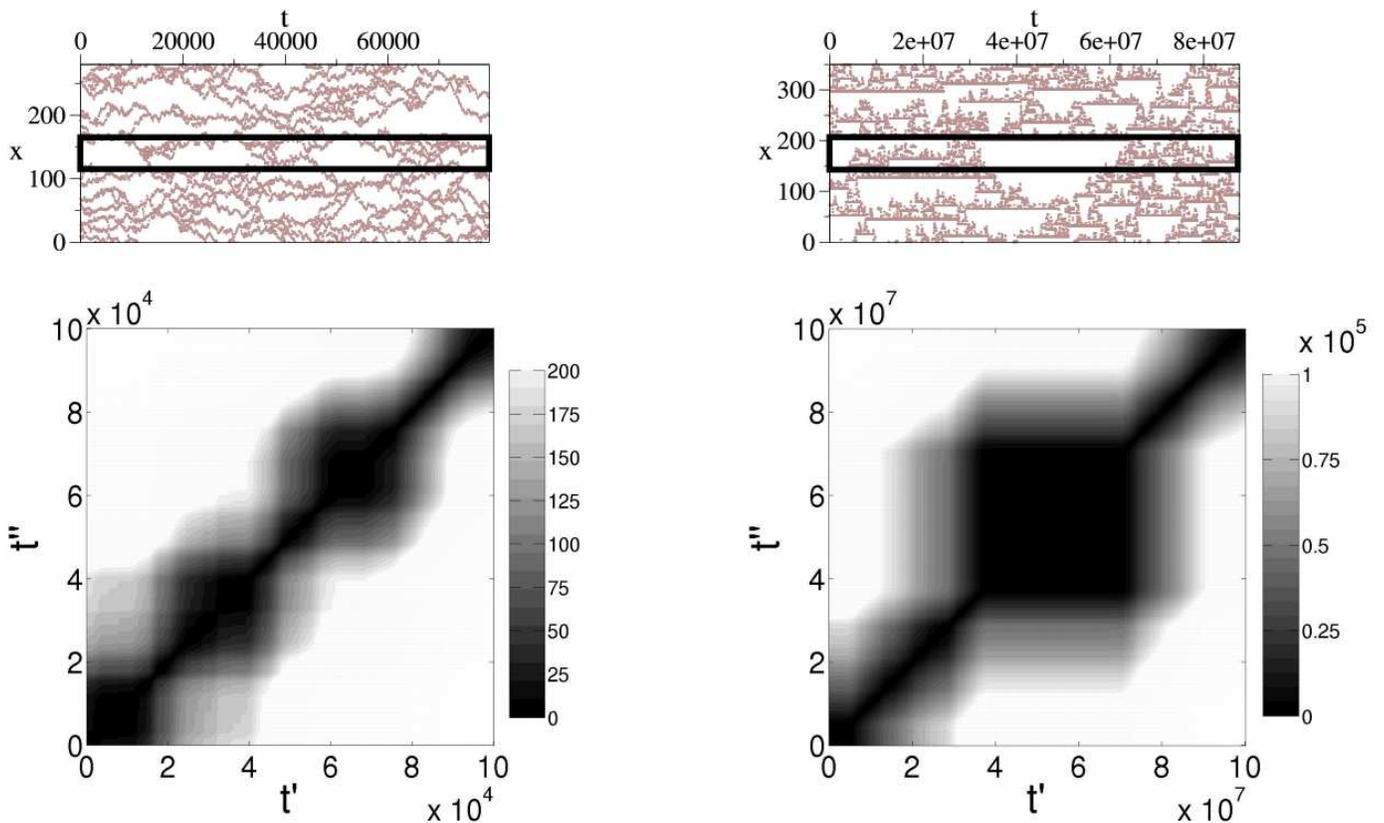}\\
  \caption{(Top) Trajectories for the FA model (left) at inverse temperature $\beta=3$, and for the East model (right) at $\beta=4$.  The length of the trajectories is several relaxation times.  The boxes indicate the sub-trajectories used to calculate the distance matrices.  The subsystem size is $N=50$ for both models. 
(Bottom) Corresponding distance matrices $\Delta^2(t',t'')$ for the sub-trajectories in the FA model (left) the East model (right).}
\label{trajectories-DM-fsm}
\end{figure*}

The top panels of Fig. \ref{trajectories-DM-fsm} show trajectories for the FA and East models.  Mobile sites are dark and immobile sites are white. In both cases the trajectory is several times the length of the structural relaxation time, $\tau_{\alpha}$, at the corresponding temperature.  The trajectories illustrate the familiar features of dynamic facilitation \cite{Garrahan-Chandler}: at low temperatures trajectories are spatially heterogeneous, excitations form continuous lines in space-time, and there are large inactive space-time ``bubbles''.

In analogy with Ref.\ \cite{Appignanesi} we study metabasin transitions by
computing the ``distance matrix'' (DM), defined as:
\begin{equation}
    \Delta^2(t',t'')=\frac{1}{N} \sum_{i=1}^N |k_i(t')-k_i(t'')|^2,
    \label{Delta}
\end{equation}
where $k_i(t)$ is the number of times site $i$ has changed state in time $t$. 
Hence, $\Delta^2(t',t'')$ represents the average squared number of configuration changes, or ``kinks'', in the time interval $(t',t'')$.  Our definition of the DM should be compared to that used in Refs.\ \cite{Appignanesi,Appignanesi2,Frechero,Binder} for atomistic models, 
\begin{equation}
     \tilde{\Delta}^2(t',t'')=\frac{1}{N} \sum_{i=1}^N |\mathbf{r}_i(t')-\mathbf{r}_i(t'')|^2,
     \label{Delta2}
\end{equation}
where $\mathbf{r}_i(t)$ is the position of particle $i$ at time $t$ \cite{Ohmine}. In this case $\tilde{\Delta}^2(t',t'')$ measures the system's averaged squared displacement between times $t'$ and $t''$. 

Just as in the atomistic case, in order to focus on individual metabasin transitions we need to consider a small enough subsystem \cite{Berthier-Garrahan}.  The boxes in the top panels of Fig.\ 1 indicate the sub-trajectories we use for the analysis.  The spatial extension of our subsystems in each case is comparable to the dynamic correlation length, $\langle n_i \rangle^{-1} \approx e^{\beta}$ \cite{Garrahan-Chandler,Berthier-Garrahan}.  Due to the heterogeneous nature of the dynamics, mobility within a sub-region changes dramatically over the course of time. For long periods the system can remain inactive until eventually a sudden burst of activity occurs as an excitation line enters and eventually passes through the observation window.  The bottom panels of Fig.\ 1 show the DMs for the chosen sub-trajectories.  

The DMs for the simple FA and East models show remarkable similarity to those seen in atomistic models in Ref.\ \cite{Appignanesi}.  The large black squares in the DMs of Fig.\ 1 correspond to times when the subsystem is trapped in a metabasin.  It is clear from inspection of the associated sub-trajectories that these periods correspond to the inactive bubbles in space-time.  These quiescent periods are  punctuated by bursts of activity that rearrange the subsystem into a different metabasin, seen as new black square in the DM. Since our simple microscopic models are built purely from the idea of facilitation it is clear that the burst of mobility during a MB transition is not the result of a democratic motion of particles but rather a sequence of cooperative events that occur on a time scale far smaller than the typical MB lifetime.

\begin{figure*}[t]
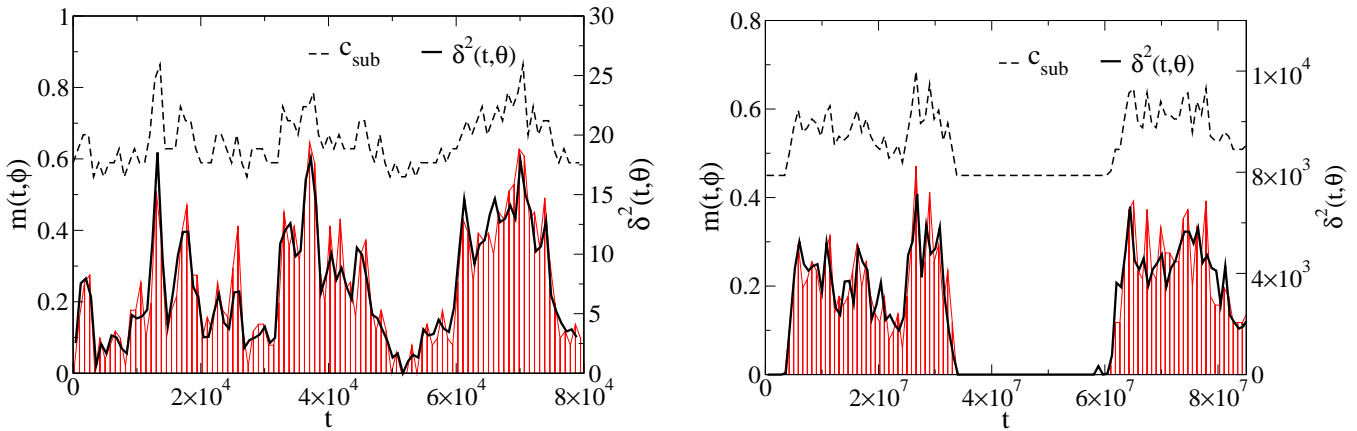

  \centering
  \includegraphics[width=1\columnwidth]{fig2a.eps}
\hspace{10pt}
  \includegraphics[width=1\columnwidth]{fig2b.eps}
  \caption{Average squared kinks $\delta^2(t,\theta)$ (solid line) for the FA model (left) and East model (right). $\theta=800$ for the FA model, and $\theta=8\times10^5$ for the East model.  Also shown is the fraction of sites $m(t,\phi)$ with a higher than average number of kinks in the time interval $\phi=200$ (FA) and $\phi=2\times10^5$ (East). Dashed lines indicate the concentration of excitations within the sub-region, $c_{\mathrm{sub}}(t)$; the curves are unscaled and have been shifted vertically to aid clarity.}
  \label{ask-fsm}
\end{figure*}

To further illustrate this idea consider the measure $\delta^2(t,\theta)$, the averaged squared kinks within the sub-region in a time interval $\theta$,

\begin{equation}\label{average-square-kinks}
    \delta^2(t,\theta)=\Delta^2(t - \theta/2, t + \theta/2) ,
\end{equation}
which is shown in Fig. \ref{ask-fsm} (solid lines).  This quantity is simply $\Delta(t',t'')$ measured along the diagonal $t''=t'+\theta$. Again, following \cite{Appignanesi} we specify that $\theta$ is significantly smaller than $\tau_{\alpha}$ but larger than the timescale associated with any microscopic motion. For the FA model we choose $\theta=800$ and for the East model $\theta=8\times10^5$. From the figure we see that $\delta^2(t,\theta)$ is low during periods of inactivity in the trajectory and there are sharp peaks signifying the transition between metabasins as an excitation line sweeps through the observed region.

The notion of democratic particle motion \cite{Appignanesi} was inferred from the fact that the peaks in the average mobility did not result from a substantial displacement of a few particles, but rather an increase in the average displacement of a significant fraction (as much as $30\%$) of particles within the subsystem.  Something similar occurs in the FSMs.  In Fig.\ 2 we also show $m(t,\phi)$, the fraction of sites in the sub-region that have experienced a higher than average number of kinks in the time interval $\phi$. The peaks in $m(t,\phi)$ coincide with those of $\delta^2(t,\theta)$ and at the corresponding time a significant fraction of sites within the sub-region exhibit a high number of kinks.  However, this apparent democratic motion is the of result of facilitation, the global relaxation event resulting from a sequence of locally facilitated ones.  The extent to which the excitation line penetrates the sub-region determines the scale of the apparent democracy.

As illustrated in the space-time trajectories of Fig. \ref{trajectories-DM-fsm} for both the FA and East models it is the presence of excitations that drives the dynamics within a given region of the system. The features observed in the DMs are a direct result of the fluctuating concentration of excitations within the sub-region, $c_{\mathrm{sub}}$. To illustrate this point we plot $c_{\mathrm{sub}}(t)$ along each trajectory, shown as the dashed line in both panels of Fig. \ref{ask-fsm} (note that the curves for $c_{\mathrm{sub}}$ are unscaled and have been shifted vertically to aid visualisation). We can see that the quantities $c_{\mathrm{sub}}(t)$ and $\delta^2(t,\theta)$ behave similarly.

\begin{figure*}[t]
  \centering
  \includegraphics[width=0.65\columnwidth]{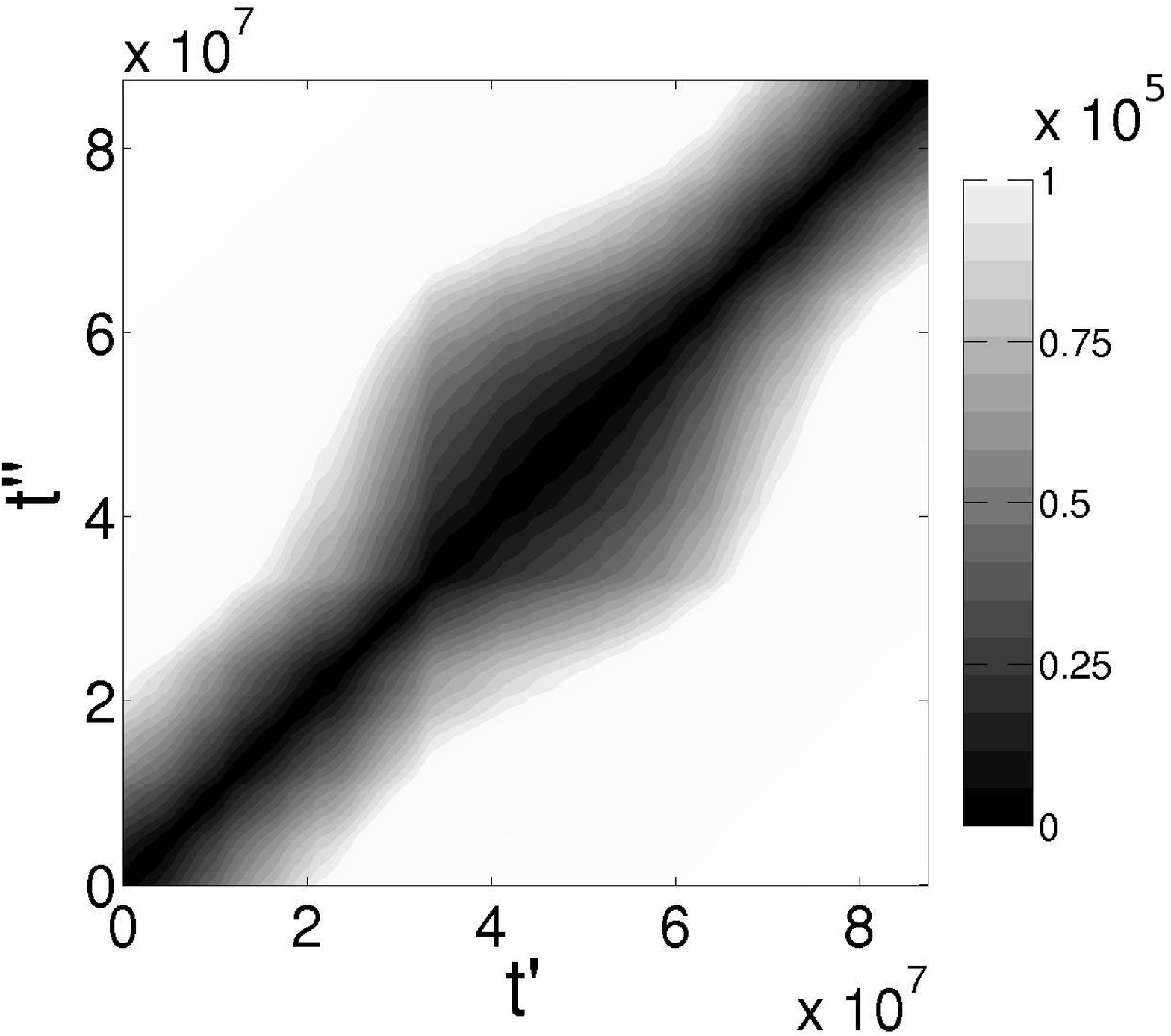}
  \includegraphics[width=0.65\columnwidth]{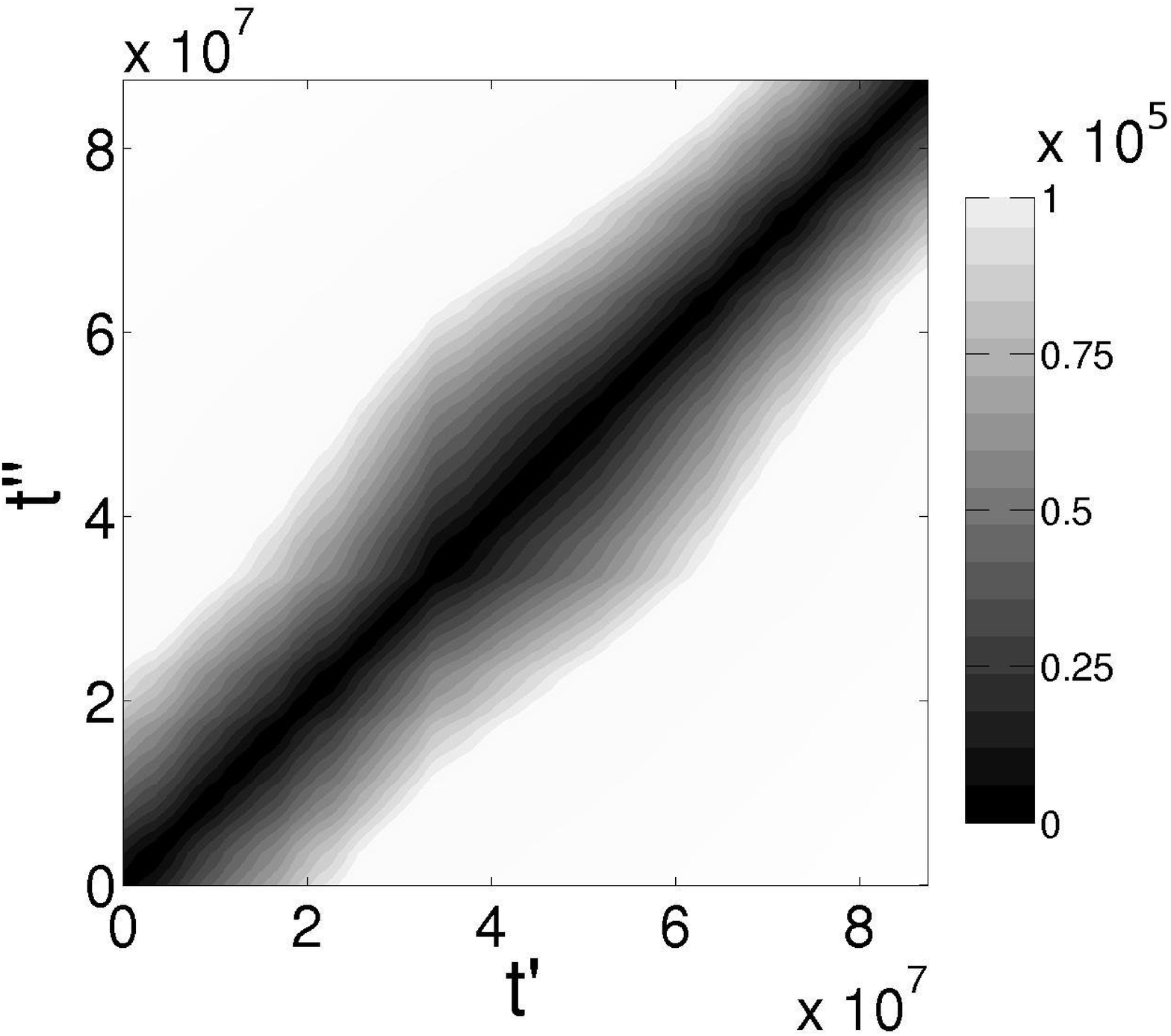}  
  \raisebox{0.2cm}{\includegraphics[width=0.75\columnwidth]{fig3c.eps}}  
  \caption{(Left and centre) Distance matrices for the East model at two different subsystem sizes, $N=150$ (left) $N=300$. (centre). (Right) Average squared kinks $\delta^2(t,\theta)$ (solid line) for the East model subsystem of size $N=300$ where $\theta=8\times10^5$.  Also shown is the fraction of sites $m(t,\phi)$ with a higher than average number of kinks in the time interval $\phi=2\times10^5$. The dashed line (which has been raised vertically and is unscaled) indicates the concentration of excitations within the sub-region, $c_{\mathrm{sub}}(t)$. The fluctuations coincide perfectly with $m(t,\phi)$.}
  \label{DMs-fsm-size}
\end{figure*}

For larger subsystems the sharp features seen in the DMs begin to disappear as the dynamics becomes more homogeneous in nature. Fig. \ref{DMs-fsm-size} shows DMs for the East model at two larger system sizes; as the size of the sub-region is increased the island structure of the DMs is quickly lost forming a dark band along the diagonal $t'' = t'$. The increasing homogeneity in the temporal variation of the subsystem dynamics is also reflected in the average squared kinks, right panel of Fig. \ref{DMs-fsm-size}. Here we show the average squared kinks for the East model using a subsystem of size $N=300$, where $\theta=8\times10^5$ once again. Although small in comparison to Fig. \ref{ask-fsm} the dynamics within the sub-region still fluctuates significantly, a detail which is somewhat obscured in the DM itself. In addition fluctuations in the fraction of sites experiencing above average kinks, $m(t,\phi)$, still coincides with $\delta^2(t,\theta)$. The dashed line again indicates the concentration of excitations within the sub-region, the fluctuations of which agree precisely with $m(t,\phi)$.

\bigskip

\section{Metabasin transitions in constrained lattice gases} 
Another class of simple KCMs are 
constrained lattice gases \cite{Kob-Andersen,Jackle-tlg,Ritort-Sollich,KA}. 
Here we consider the two-vacancy assisted triangular lattice gas, or (2)-TLG
\cite{Jackle,Pan}.  The system consists of hard-core particles that move on a two-dimensional lattice of triangular geometry; there are no static correlations between particles and at most each site can hold one particle at a time. Any particle on the lattice can only move to one of its six nearest neighbour sites if the following rules are satisfied: (i) the target site is unoccupied and (ii) both the two mutual nearest neighbours of the initial and target site are also empty. The physical interpretation of the dynamical rule is the steric constraint on particle motion that occurs within a dense fluid. The dynamics of the model is highly collective, for a particle to be able to move first requires the cooperative rearrangement of many of its neighbours.  

We again choose to examine the dynamics within a small sub-region of a large system using the definition of the distance matrix given in Eq.\ (\ref{Delta}). This definition is convenient since it accounts for all particles which enter or leave the sub-region during the course of a given trajectory, rather than those simply present at a given time. Here a kink corresponds to a particle entering or leaving a site. By using a sub-system within an extensive lattice we also avoid the potential problem of forming a backbone of frozen particles that could percolate the region of interest \cite{Jackle-tlg,Kob-Andersen}.  

\begin{figure}[b]
  \centering
  \includegraphics[width=\columnwidth]{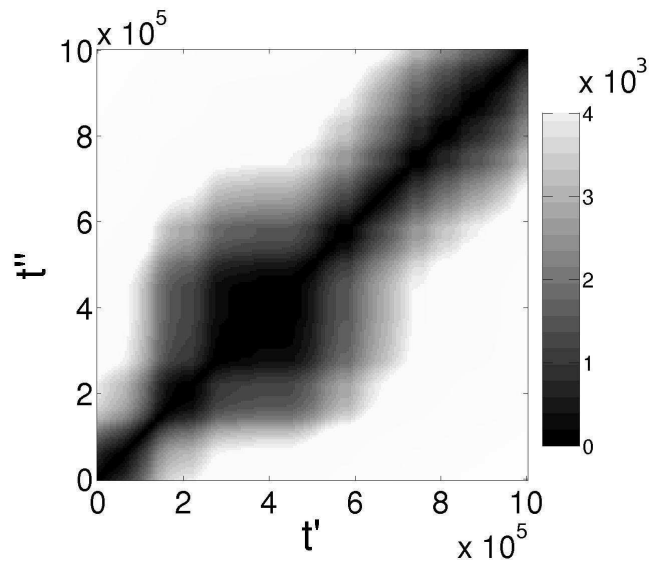}
  \caption{Distance matrix $\Delta^2(t',t'')$ for the (2)-TLG. The data corresponds to a $10 \times 10$ subregion from a lattice at particle density $\rho=0.79$ (see Fig.\ref{ask-snapshot-tlg}).}
\label{DM-tlg}
\end{figure}

\begin{figure*}[t]
  \centering
  \includegraphics[width=\columnwidth]{fig5a.eps}
  \raisebox{0.5cm}{\includegraphics[width=\columnwidth]{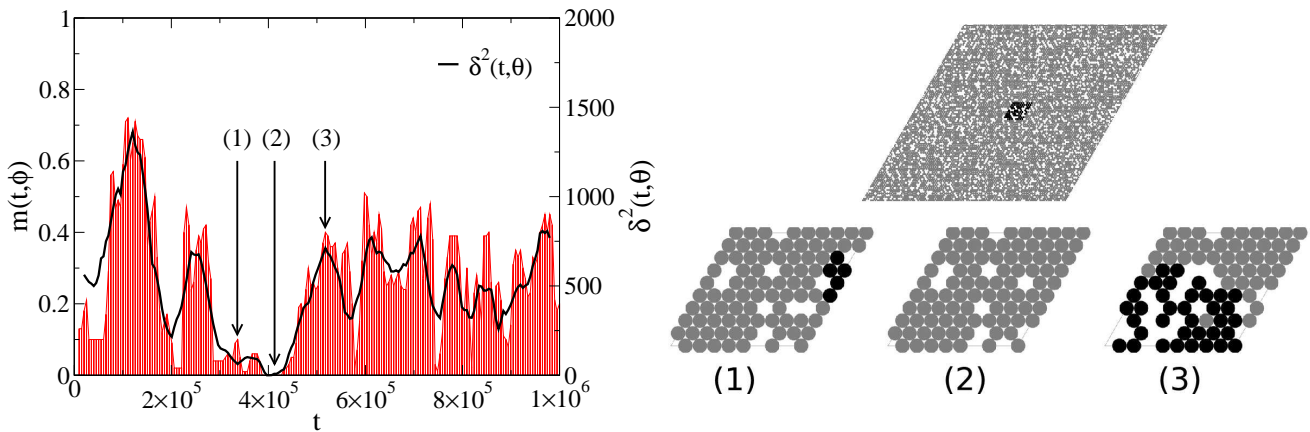}}
  \caption{(Left) Average squared kinks $\delta^2(t,\theta)$ (solid line) for the (2)-TLG model.  Also shown is the fraction of sites $m(t,\phi)$ with a higher than average number of kinks in the time interval $\phi=10^4$.  (Right) The top figure shows the sub-region considered. The dynamics within this sub-system make up the distance matrix shown in Fig.\ 3 and in the left panel. The numbered images are snapshots of the particle field within the sub-region at the times indicated in the left panel. Particles which have moved position in the time interval $\phi=10^4$ are coloured black.}
\label{ask-snapshot-tlg}
\end{figure*}

Fig. \ref{DM-tlg} shows an example distance matrix for a $10\times10$ sub-region taken from a (2)-TLG lattice at particle density $\rho=0.79$. The total simulation length is $10^6$, the persistence time at the chosen density \cite{Pan}. The DM exhibits the same features found in the FSMs and in atomistic models. The trajectory is characterised by extended regions of low activity, which appear as dark squares in the DM,  with relatively quick transitions between them \cite{Movie}. Fig. \ref{ask-snapshot-tlg} (left) shows the average squared kinks, $\delta^2(t,\theta)$, for the same trajectory, where $\theta=4\times10^4$. Also shown is the fraction of sites with above average kinks in a time window $\phi = 10^4$. As for the FSMs, at the peaks in $\delta^2(t,\theta)$ a significant fraction of sites in the sub-region are experiencing a high level of activity. From direct analysis of the particle trajectories we find the MB-like regions correspond to configurations in which the majority of particles within the sub-region are blocked and remain frozen in position for large periods of time. A burst of motion results from a sequence of unlocking events which enables the particles to quickly rearrange themselves. After the effective mobility excitation passes the particles are left in a completely or partially frozen configuration. This dynamical picture is in direct agreement with previous analysis of the TLG models \cite{Jackle-tlg,Pan,Lester}.  

\begin{figure*}[t]
  \centering
  \includegraphics[width=\columnwidth]{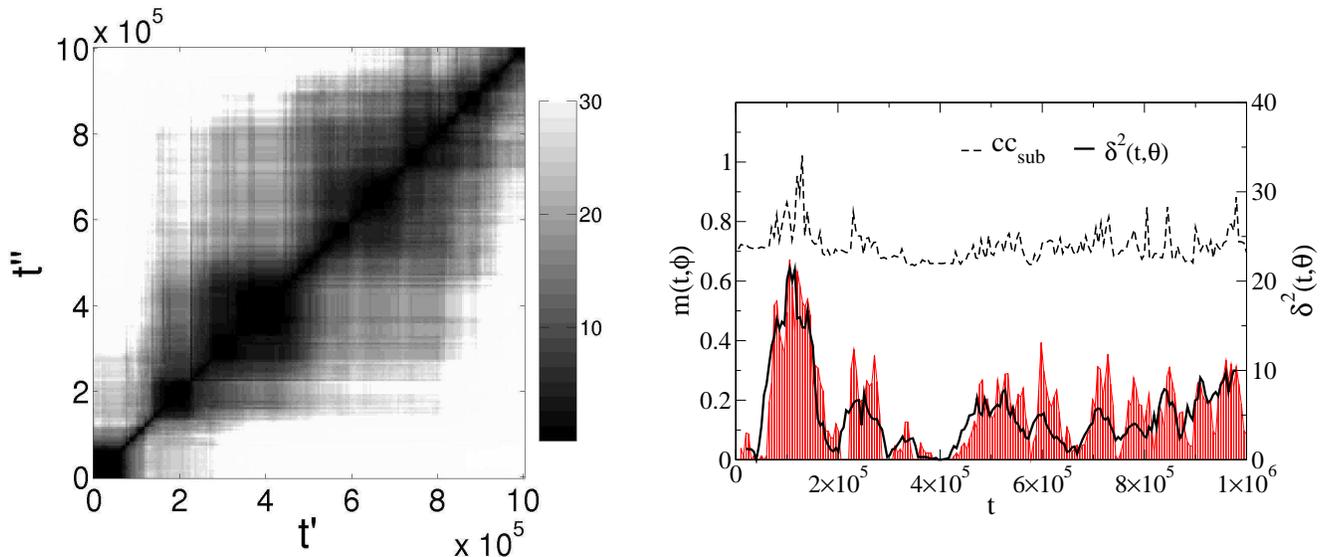}
  \raisebox{0.7cm}{\includegraphics[width=\columnwidth]{fig6b.eps}}
  \caption{(Left) Distance matrix for the same (2)-TLG system of Fig. \ref{ask-snapshot-tlg} now using the distance matrix as defined in Eq. (\ref{Delta2}). In comparison with Fig. \ref{DM-tlg} it is evident both approaches capture the same dynamical information. (Right) This is also apparent in a plot of the average squared displacement $\delta^2(t,\theta)$ (solid line) which agrees closely with the average squared kinks of Fig. \ref{ask-snapshot-tlg}. The same value of $\theta$ has been used in both figures. Also shown is the fraction of particles $m(t,\phi)$ with a higher than average displacement in the time interval $\phi=10^4$. The peaks in $\delta^2(t,\theta)$ are clearly accompanied by a significant displacement of a large fraction of the particles within the subsystem. The dashed line indicates the average cluster connectivity (see text) of particles within the sub-region. The curve is unscaled and has been raised vertically to aid clarity.}
\label{DM-asd-tlg}
\end{figure*}

Due to the particle nature of the TLG model it is also possible to consider the original definition of the DM as given in Eq. (\ref{Delta2}), i.e. the subsystem's averaged squared displacement between times $t'$ and $t''$. Since the problem of a backbone prevents simulation of a small system it is necessary to modify the summation such that only those particles present in the sub-region at time $t'$ are considered. Fig. \ref{DM-asd-tlg} shows (left panel) the resulting DM for the same $10 \times 10$ subsystem used previously. The DM is similar to that of Fig. \ref{DM-tlg} indicating that both definitions of the distance matrix capture the same dynamical information. This is also evident when analysing the average squared displacement $\delta^2(t,\theta)$ (again using $\theta=4\times10^4$) as shown in the right hand panel (solid line). Also shown is the fraction of particles with above average displacement in the time interval $\phi = 10^4$. In agreement with the results of atomistic simulations the peaks in the average squared displacement are accompanied by a large displacement of a significant fraction of the particles within the subsystem.

In a recent paper \cite{Lester} we identify a possible elementary excitation for the (2)-TLG by studying the dynamics of the model with the iso-configurational (IC) method \cite{Harrowell2}. Here one constructs an ensemble of equal length trajectories which share a common initial particle configuration but with different initial particle momenta (for the lattice gas Monte Carlo used to simulate the TLG models one uses a different sequence of attempted moves for each trajectory). As for the Lennard-Jones liquids studied previously \cite{Harrowell2}, we found \cite{Lester} heterogeneity in the spatial distribution of the dynamic propensity, $\left\langle \Delta\mathbf{r}_i^2\right\rangle_{\mathrm{IC}}$, the squared displacement of particle $i$ averaged over all trajectories within the IC ensemble. This heterogeneity was found to correlate well with a structural measure, the connectedness of particles to extended clusters of vacancies \cite{Lester}.  This was defined as the total sum of the vacancy cluster sizes to which a particle is connected to through its nearest neighbours \cite{Lester}. Cluster connectivity was also found to provide a good prediction of the instantaneous particle mobility within a single trajectory. The most mobile particles at any given time are on average the most connected, typically forming a ring around a large cluster of vacancies. To illustrate the role of cluster connectivity for the dynamics of the (2)-TLG we plot the average cluster connectivity of particles within the sub-region along the trajectory, shown as a dashed line in the right hand panel of Fig. \ref{DM-asd-tlg}. Once again the curve is unscaled and has been shifted vertically for clarity. The fluctuations in the cluster connectivity match well (in particular the highest peaks) with those of the average squared displacement (or average squared kinks) further illustrating the role of high connectivity particles as the relevant dynamic excitations for the (2)-TLG model. 

\bigskip

\section{Conclusions} 
We have shown that transitions between metabasins in glass formers can be understood in terms of dynamic facilitation.  The apparent ``democratic'' collective particle rearrangements are really a sequence of localised and facilitated events.  Observed on small enough lengthscales they just appear sudden and homogeneous.  They are in essence another manifestation of dynamic heterogeneity. 

The distance matrix, Eq.\ (\ref{Delta}), encodes the trajectories in an elegant way and the similarity between the DMs we observe here in KCMs, Figs. \ref{trajectories-DM-fsm} and \ref{DM-tlg}, and the ones measured in molecular dynamics simulations \cite{Appignanesi,Appignanesi2,Frechero} is remarkable.  For KCMs we know that the observed structure in the DM is just a projection of the space-time correlations in trajectories, that is, of inactive bubbles bounded by excitations lines (itself a consequence of phase coexistence between active and inactive dynamical phases in KCMs \cite{ruelle}).  Our results here give further support to the view that atomistic models have trajectories with similar features.

\bigskip
We thank Gustavo Appignanesi for discussions.  
This work was supported by EPSRC grant GR/S54074/01 and University of Nottingham grant FEF3024.

\end{document}